\documentclass[a4paper,twocolumn,aps,pra,showpacs]{revtex4}

\usepackage{graphicx}
\usepackage{amsmath}
\usepackage{amstext}

\def\diag{\mathop{\mbox{diag}}\nolimits}
\def\eig{\mathop{\mbox{eig}}\nolimits}

\begin{document}
\title{Quantum homogenization and state randomization in semi-quantal spin systems}

\author{M\'aty\'as Koniorczyk}
\email{kmatyas@gamma.ttk.pte.hu}
\affiliation{Institute of Physics, University of P\'ecs, H-7624 P\'ecs, Ifj\'us\'ag \'utja 6, Hungary}
\affiliation{Research Institute for Solid State Physics and Optics, Hungarian Academy of Sciences, H-1525 Budapest, P.O. Box 49., Hungary}
\author{\'Arp\'ad Varga}
\affiliation{Institute of Physics, University of P\'ecs, H-7624
  P\'ecs, Ifj\'us\'ag \'utja 6, Hungary}
\author{Peter Rap\v can}
\affiliation{Research Center for Quantum Information,
Institute of Physics, Slovak Academy of Sciences
84511 D\' ubravsk\' a cesta 9, Bratislava Slovakia}
\author{Vladim\' \i r Bu\v zek}
\affiliation{Research Center for Quantum Information,
Institute of Physics, Slovak Academy of Sciences
84511 D\' ubravsk\' a cesta 9, Bratislava Slovakia}

\date{\today}
\begin{abstract}
  We investigate dynamics of semi-quantal spin systems in which
  quantum bits are attached to classically and possibly stochastically
  moving classical particles. The interaction between the quantum bits
  takes place when the respective classical particles get close to
  each other in space. We find that with Heisenberg XX couplings
  quantum homogenization takes place after a time long enough,
  regardless of the details of the underlying classical dynamics. This
  is accompanied by the development of a stationary bipartite
  entanglement. If the information on the details of the motion of a
  stochastic classical system is disregarded, the stationary state of
  the whole quantum subsystem is found to be a complete mixture in the
  studied cases, though the transients depend on the properties of the
  classical motion.
\end{abstract}
\maketitle

\section{Introduction}

Quantum mechanics provides the best known description of microscopic
physical systems. In most of the cases, however, the system to be
described is a part of a larger system, its environment, and these
interact. As a result there are correlations, usually both of
classical and quantum nature, developing between them.  From the
system's point of view this results in a modification of the behavior
which usually leads to the loss of the very quantum mechanical
features. This phenomenon is referred to as a decoherence.

The study of decoherence leads to the understanding of mesoscopic and
macroscopic systems, that is, the emergence of classical behavior.  In
addition, decoherence constitutes the main obstacle of exploiting the
quantum mechanical nature of system in applications such as quantum
information processing. The mathematical description of decoherence
aims at the derivation of the system's dynamics without the detailed
description of its environment. In many of the cases this is done
based on the full description of the system and its environment, which
are realistic physical systems, by applying several approximations. In
quantum optics, for instance, the decoherence of a single mode of the
electromagnetic field is described by taking into account the
interaction of the infinite set of other field modes by perturbation
theory and the Born and Markov approximations. The result is a
Lindblad master equation for the density operator of the system.

Besides of the study of "real" physical systems, the introduction of
simplified models is of some use. These are the microscopic models of
decoherence. In this case the aim is to give a detailed understanding
of the process of decoherence via a more detailed description of the
system and the environment, which is feasible either analytically or
numerically. Some of such models address the decoherence of a one or a
few two-level systems (quantum bits) in the presence of additional
quantum bits, that is, a spin-bath environment.  Besides its
simplicity, this choice is motivated by the relevance in quantum
computing and the relation to real-world scenarios such as two-level
atoms or solid-state systems.  Such models bear experimental relevance
in some cases and many of them can be implemented with cold bosons on
optical lattices, as pointed out by Rossini et
al.~\cite{RossiniCGMF07} recently.

There are several analytically solvable and physically realistic models
of spin baths, such as e.g.~the Tessieri-Wilkie
model~\cite{TessieriW03}, in which the role of the interactions
amongst the quantum bits of the environment is apparent.  The issue of the
interaction between the environment spins is studied in detail by
Dawson et al.~\cite{DawsonHMM05}, who point out the role of the
monogamy of entanglement as described by the Coffman-Kundu-Wootters
inequalities~\cite{PhysRevA.61.052306, osborne:220503} in the
decoherence features.  In a very recent publication Camalet and
Chitra~\cite{CamaletC07} have studied the decoherence of a qubit due
to the presence of random interactions with a spin bath.  They point
out the non-Markovian behavior at low temperatures as well as the
role of the intra-environment interactions.  Pi\~ neda et
al.~\cite{PinedaGS07} have studied the decoherence of two
non-interacting qubits in the framework of random matrix theory in
very detail.

An important class of microscopic models is that of quantum
homogenization~\cite{ZimanSBHSG02,ScaraniZSGB02}. An environment
consisting of many qubits is considered, each in the same
single-particle state $\varrho $ (Though the global state of the
environment may be a pure one assuming the presence of initial
correlations in the environment.).  The system quantum bit is
initially in an arbitrary state $\sigma $.  The interaction between
the quantum bits is defined by a two-qubit quantum gate, the partial
swap gate generated by a Heisenberg-XXX (that is, the isotropic
Heisenberg) interaction.  In its simplest formulation quantum
homogenization is a collision-type model: the system qubit interacts
with the reservoir qubits one-by-one. (Note that a collision-based
model with a qubit as a system and a classical field as a reservoir
was used by Di\'osi et al.~\cite{DiosiFK06} recently to reveal the
relation between thermodynamical and information theoretic entropy.)
The aim of the model is to demonstrate that in this framework the
state of the system quantum bit and each reservoir quantum bit will
evolve infinitesimally close to each other and also to the initial
state $\varrho$ of the reservoir.  This process is rather similar to
the thermalization in classical thermodynamics: the temperature of a
small system will become close to the original temperature of a large
environment, while this latter does not change significantly. For
qubits the single-qubit density operator plays the role of the
"temperature", and qualitatively the same situation arises. The
evolution of the system qubit is Markovian described by a discrete
semigroup which is a stroboscopic image of a continuous-time Markovian
evolution. It is also shown that a distributed pairwise entanglement
arises in the system which tends to a stationary value and saturates
the Coffman-Kundu-Wootters inequalities~\cite{ZimanSB03}. The process
is basis independent (or covariant) due to the nature of the partial
swap operation, thus it works for arbitrary initial system and
reservoir states. The partial swap is the only gate with this
property.

The model of quantum homogenization was further generalized to include
pairwise interactions within the reservoir. Assuming that at any time
step a randomly chosen pair of qubits (system or reservoir) can
interact, one obtains a Markovian evolution of the whole system again.
However due to the intra-environment interactions, the evolution of
the single qubits is not Markovian anymore. It was found that for an
actual sequence of interactions, the state of each qubit will
fluctuate around the one obtained by the collision-based
homogenization. The fluctuations appear to decrease with the growing
size of environment. If one simulates many actual evolutions and
constructs an effective density operator for each time step, this will
show ``genuine'' homogenization.  The same holds for time-averaged
states.  The model with a very small (i.e. two-qubit) reservoir was
recently studied by Benenti and Palma~\cite{BenentiP07} who have found
that irreversibility can emerge even in this case, after
time-averaging.

The stochastic homogenization model is a special case of a more
general scenario termed as a semi-quantal spin gas model. In this case
a set of classical particles (e.g. classical atoms) is considered to
move according to a certain model, e.g. an ideal Boltzmann gas. Each
particle has an internal degree of freedom which is considered to be
quantum mechanical. This can be a half spin or two possible hyperfine
states, termed as a quantum bit in what follows. The quantum
mechanical part of the system is thus multipartite, consisting of the
quantum bits attached to the respective classical particles. This
essentially models any system in which quantum bits are attached to
addressable and distinguishable entities which are allowed to move in
space for some reason.  The state of the quantum system has no effect
on the classical motion whatsoever. On the other hand, if some of the
particles fulfill a certain \emph{collision condition}, e.g. they
collide in the case of the Boltzmann gas model, a prescribed quantum
gate acts on the respective qubits.

Semi-quantal spin gases were studied by Briegel et
al.~\cite{HartmannCDB05}. Besides the Boltzmann gas as an underlying
classical model, they have considered a lattice gas with on-site
exclusion, which shows a spatially correlated behavior. In this case
the particles are located on a discrete lattice and they can hop to
the neighboring empty sites stochastically. The collision condition
is that the two particles should be at neighboring sites. In
Ref.~\cite{HartmannCDB05} an Ising coupling is considered between the
respective quantum bits.  For a particular evolution of the microstate
of the classical system one can calculate the evolution of the quantum
mechanical one, given its initial state. In the case of the Boltzmann
gas, however, only the macrostate of the classical system is known,
while the lattice gas moves stochastically by nature. Thus given a
pure quantum state as initial condition, one obtains a pure state
depending on time which has random parameters. Due to the
advantageous properties of the pairwise Ising couplings (e.g. they
commute), for certain initial states it is possible to calculate the
density operator of an arbitrary subsystem of the rather big quantum
mechanical system. In Ref.~\cite{HartmannCDB05} the authors study
various aspect of the arising non-Markovian evolution including
decoherence and entanglement behavior.

Motivated by the above summarized results we intend to consider
further microscopic models of decoherence and address additional
issues. In particular we will consider the following classical
models:
\begin{itemize}
\item completely random pairwise interactions,
\item classical particles moving in a three-dimensional box, colliding
  elastically with each other and the wall (the ``billiard ball''
  model), the qubits interact if the respective particles collide,
\item discrete sites along a line, on which particles can hop to the
  neighboring empty sites (one-dimensional lattice gas), the qubits
  interact if the respective particles are next to each other.
\end{itemize}
Our one-dimensional lattice gas can be imagined as a spin chain with
vacancies where the classical particles carrying the spins can jump to
a vacancy. Since a similar dynamics is present in any solid for
thermodynamical stability reasons, we believe that this model might
bear some experimental relevance. From the point of view of
decoherence models, we are able to describe a very strongly
self-interacting reservoir. We will consider the Heisenberg-XX and
the Ising Hamiltonians as well. We perform computer simulations of the
system to deduce our results.

First we will consider a given particular evolution of the classical
system and Heisenberg-XX type couplings. This choice of the
Hamiltonian is motivated by the previous studies of quantum
homogenization. In the case of the billiard ball model the classical
trajectory is uniquely determined by the initial conditions for the
classical system while for stochastic dynamics it is generated by
random deviates.  We seek for quantum homogenization and find that it
indeed appears for all the classical dynamics after a time long
enough. The main conclusion will be that the characteristics of the
classical evolution do not play a relevant role in the
homogenization, that is, in the state of a single-qubit subsystem. We
also show that there is a distributed bipartite entanglement in the
system.

Another possible attitude in the case of the stochastic evolutions is
that one ignores the information on the actual classical motion. This
loss of information results in the increasing entropy of the state of
the quantum mechanical subsystem. We will study the so-arising
decoherence in detail, both for Ising and Heisenberg-XX couplings. We
show that it leads to a completely mixed state in the accessible part
of the Hilbert space rather quickly, for all the considered models.

This paper is organized as follows: in Section~\ref{sect:trajs} we
consider individual evolutions and study the behavior of quantum
homogenization. In Section~\ref{sect:globalstate} we consider the
dynamics of the effective density matrix of the whole quantum
subsystem, built up as the convex combination of the system's state in
case of different particular evolutions. In Section~\ref{sect:concl}
the results are summarized and the conclusions are drawn.

\section{Individual trajectories: from spin-chains to homogenization
models}
\label{sect:trajs}

In this Section we seek for generalizations of the quantum
homogenization models. In the already studied versions of these, a set
of qubits is considered, one of which plays the role of the system
while the others play the role of the environment.
The environment qubits are initially in a state $\varrho$, while the
system qubit is in another arbitrary state $\sigma$. The interactions
are assumed to be bipartite: in each time step a chosen pair of two
quantum bits interacts via a partial swap quantum gate which can be
written, up to an irrelevant phase factor, as
\begin{equation}
  \label{eq:pswap}
U_{\text{pswap}}(\eta)=\exp \left( -i \eta \hat H^{(\text{XXX})} \right),
\end{equation}
where
\begin{equation}
\quad H^{(\text{XXX})}=
\sigma_x \otimes \sigma_x + \sigma_y \otimes \sigma_y + \sigma_z \otimes \sigma_z
\end{equation}
is the Heisenberg XXX interaction. Depending on the model the
reservoir qubit interacts with the reservoir particles one-by-one
once, or randomly chosen pairs will interact.  Finally, all the
single-qubit density operators, describing either the state of the
system qubit or any of the reservoir qubits will be approximately
equal: the state of the system qubit will be ``diluted'' in its
environment.

Instead of $H^{(\text{XXX})}$ let us consider a pairwise interaction
described by the Heisenberg XX Hamiltonian
\begin{equation}
  \label{eq:Hxx}
  H^{(\text{XX})}= \sigma_x \otimes \sigma_x + \sigma_y \otimes \sigma_y.
\end{equation}
The motivation for this choice is twofold. First, these kind of
couplings play an important role in quantum state transfer in spin
systems~\cite{christandl:187902}. Secondly, the evolution generated by
this pairwise Hamiltonian between any configuration of pairs has an
invariant subspace spanned by the vectors
\begin{equation}
  \label{eq:subspacebasis}
  |\underline{k}\rangle=|0_1 0_2 \ldots 0_{k-1}1_k0_{k+1}\ldots 0_N\rangle,
\end{equation}
thus with the initial state in Eq.~\eqref{eq:psiinsubs_normalbase},
that is, $|\underline{1}\rangle$, the state of the system at time $t$
will be a superposition of the vectors in Eq.~\eqref{eq:subspacebasis}
with different values of $k$.  (Throughout this paper we will denote
the elements of the computational basis on the Hilbert-space of a
qubit by $|0\rangle$ and $|1\rangle$, with the convention
$\sigma_z=|0\rangle\langle0|-|1\rangle\langle 1 \rangle$.) In this
basis the Hamiltonian in Eq.~\eqref{eq:Hxx} has a rather intuitive
form. If the $k$-th and $l$-th qubit is coupled, the only nonzero
matrix elements are
$\langle\underline{k}|H^{(\text{XX})}|\underline{l}\rangle=
\langle\underline{l}|H^{(\text{XX})}|\underline{k}\rangle=2$.  We
assume for the moment that in each time step a randomly chosen pair of
quantum bits interact via this coupling, thus we apply the gate
\begin{equation}
  \label{eq:xxgate}
U_{XX}(\eta)=\exp \left( -i \eta \hat H^{(\text{XX})} \right).
\end{equation}
As an initial state let us consider
\begin{equation}
  \label{eq:psiinsubs_normalbase}
  |\Psi(t=0)\rangle=|\underline{1}\rangle.
\end{equation}
where the first qubit is considered to model the system, while the
others constitute the reservoir.

If the state of the system is within the subspace in argument,
described by the density matrix
$\varrho_{\underline{k},\underline{k}'}$ expressed in the basis in
Eq.~\eqref{eq:subspacebasis}, the density operator of the $k$-th
particle is diagonal in the computational basis and it reads
\begin{equation}
  \label{eq:rhok}
  \varrho^{(k)}=
  \diag(1-p_k,p_k),\ \mbox{where}\ p_k=\varrho_{\underline{k},\underline{k}}.
\end{equation}
Here $p_k$ is the probability of finding the $k$-th qubit in the state
$|1\rangle$ in a projective measurement in the computational basis.
Another quantity to consider is the concurrence, which is used
prevalently to quantify the entanglement of two quantum bits:
\begin{widetext}
\begin{equation}
  \label{eq:concurrence}
  C_{\underline{k},\underline{k}'}=\min (0,\lambda_1-\lambda_2-\lambda_3-\lambda_3),\
  \lambda_i=\eig ( \sqrt{ \sqrt{\varrho^{(k,k')}} \tilde \varrho^{(k,k')} \sqrt{\varrho^{(k,k')}}}),
\end{equation}
\end{widetext}
where $\tilde \varrho^{(k,k')}=\sigma_y \otimes \sigma_y
\varrho^{(k,k')\ast}\sigma_y \otimes \sigma_y$ is the Wootters-tilde
and $\varrho^{(k,k')\ast}$ is the transpose of $\varrho^{(k,k')}$, the
density matrix of qubits $k$ and $k'$, in the computational basis.  In
the subspace spanned by the vectors in Eq.~\eqref{eq:subspacebasis}
this is simply twice the modulus of the coherences (off-diagonal
matrix elements) of the system's density matrix,
\begin{equation}
  \label{eq:C}
  C_{\underline{k},\underline{k}'}=2|\varrho_{\underline{k},\underline{k}'}|,
\end{equation}
as it can be verified by calculating the bipartite density matrix and
substituting it to Eq.~\eqref{eq:concurrence}.

\begin{figure}
  \centering
  \includegraphics{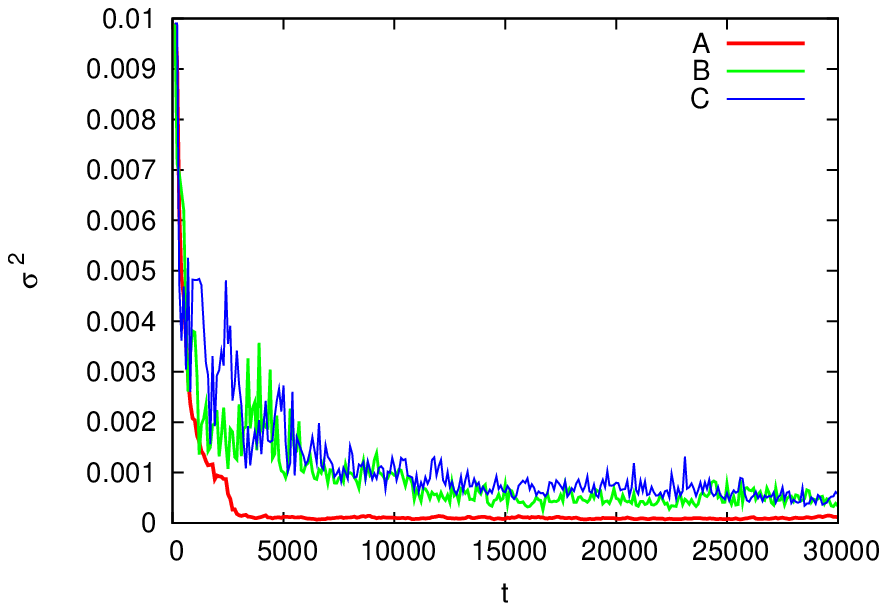}

  \includegraphics{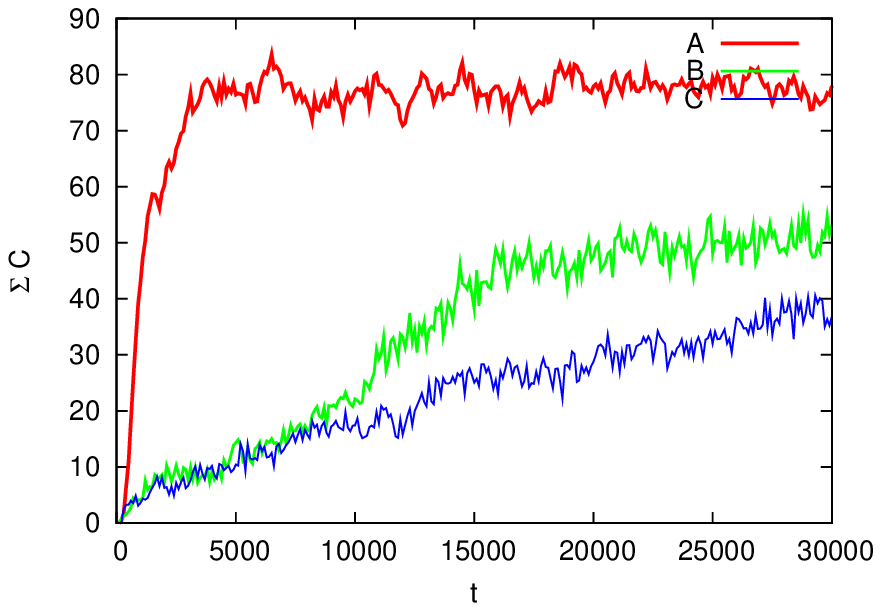}
  \caption{(color online) Evolution of the degree of inhomogeneity in
    Eq.~\eqref{eq:sigmasq} (upper figure) and the sum of the pairwise
    concurrences in Eq.~\eqref{eq:concsum} (lower figure) for
    different scenarios. A: random pairwise interactions, B:
    one-dimensional diffusive gas with 150 sites, C: one-dimensional
    diffusive gas with 200 sites. Only the values at every 100th time
    step are plotted, and the data are connected with lines to guide
    the eye. There are $N=100$ particles considered, the interaction
    strength parameter in Eq.~\eqref{eq:xxgate} and
    Eq.~\eqref{eq:xxchainu} is $\eta=0.1$.  Repeated calculations of
    the stochastic models give figures with different details but of
    the same character.}
  \label{fig:twotraj1}
\end{figure}

Let us return to the evolution $|\Psi(t)\rangle$ with the initial
condition $|\Psi(t=0)\rangle=|\underline{1}\rangle$, under the
application of the quantum gate in Eq.~\eqref{eq:xxgate} on a randomly
chosen pair of qubits in each step. As a degree of inhomogeneity of the
system we consider the square of the standard deviation of the $p_k$-s
in Eq.~\eqref{eq:rhok},
\begin{equation}
  \label{eq:sigmasq}
  \sigma^2(t)=\langle p_k^2\rangle_k-
  \langle p_k \rangle_k^2,
\end{equation}
where $\langle \ldots \rangle_k$ stands for averaging over the
particles. The smaller it is, the more homogeneous the system will
become from the point of view of single particles.  This is plotted in
Fig.~\ref{fig:twotraj1} (Upper figure, curve A). Though the
evolution shows some stochastic features due to the nature of the
interaction sequence, after a time long enough the system becomes
homogeneous. The mechanism of the homogenization is similar to the
already studied models of stochastic homogenization.

It is also interesting to consider the quantum correlations generated
during the evolution. In Refs.~\cite{PhysRevA.61.052306,KoniorczykRB05} it was shown that
the states which the initial state $|\underline{1}\rangle$ evolve into
during the dynamics generated by Heisenberg XX couplings have
bipartite entanglement only. Let us quantify this in terms of
the sum of the concurrences
\begin{equation}
  \label{eq:concsum}
  C_{\text{tot}}(t)=\sum_{k<k'} C_{\underline{k},\underline{k}'}(t),
\end{equation}
which therefore provides full information on the net amount of quantum
correlations.  As it can be seen in Fig.~\ref{fig:twotraj1} (Lower
figure, curve A), this grows and then fluctuates around a stationary
value. This is how the decoherence of single qubits arises in this
system.

\subsection*{Spin chains with vacancies}

\begin{figure}
  \centering
  \includegraphics{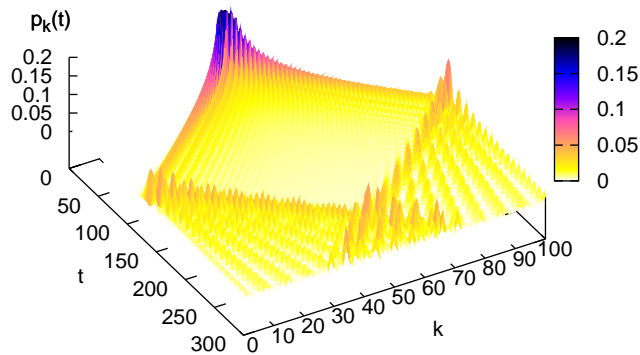}
  \caption{(color online) ``Propagation of the disturbance'' in a
    usual Heisenberg $XX$ chain of 100 quantum bits, with periodic
    boundary conditions.  Initially the qubit in the middle is in the
    state $|1\rangle$ while all the others are in the state
    $|0\rangle$ (that is, initial state is
    $|\underline{50}\rangle$). The probability $p_k(t)$ in
    Eq.~\eqref{eq:rhok} of finding the $k$-th particle in the state
    $|1\rangle$ is plotted. The values are truncated at $p_k(t)=0.2$
    to make the detailed structure more visible. The interaction
    strength in Eq.~\eqref{eq:xxchainu} is $\eta=0.1$. Notice the
    appearance of edge effects around the 100-th time step.}
  \label{fig:xxprop}
\end{figure}

Let us consider spins aligned along a chain with periodic boundary
conditions. Let the interaction be defined at the Hamiltonian level as
the sum of nearest-neighbor XX interactions, with periodic boundary
conditions as usual in statistical physics:
\begin{equation}
  \label{eq:xxchain}
  H^{(\text{chain})}=\sum_{k}H^{(\text{XX})}_{k,k+1}.
\end{equation}
Consider the dynamics generated by this Hamiltonian. To be comparable with the discrete-time models, we consider the repeated application of
\begin{equation}
  \label{eq:xxchainu}
U(\eta)=\exp \left( -i \eta H^{(\text{chain})} \right)
\end{equation}
to the initial state which is again considered to be
$|\underline{1}\rangle$. This results in a kind of ``propagation of the
disturbance'', as it can be seen in Fig.~\ref{fig:xxprop}, where the
$p_k(t)$-s of Eq.~\eqref{eq:rhok} are plotted against $k$ and $t$.
As the qubits are aligned to a chain in this case, the value of $k$
describes a kind of spatial position of the qubit in the chain.  The
propagation is a kind of broadening, but it shows interference stripes
reflecting the quantum mechanical nature of the system.

\begin{figure}
  \centering
  \includegraphics{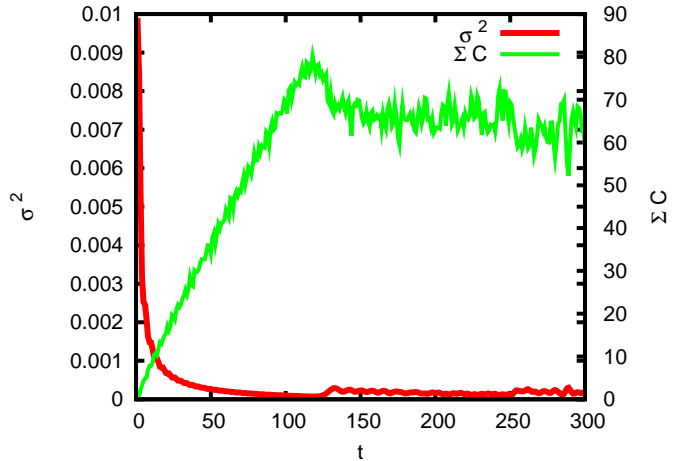}
  \caption{(color online) Evolution of the degree of inhomogeneity in
    Eq.~\eqref{eq:sigmasq} and the sum of the pairwise
    concurrences in Eq.~\eqref{eq:concsum} for the same situation as
    in Fig.~\ref{fig:xxprop}, for a Heisenberg-XX chain.}
  \label{fig:twotraj2}
\end{figure}

The question arises if a kind of homogenization effect appears even in
this very coherent scenario.  Therefore we consider again the time
evolution of the degree of inhomogeneity in Eq.~\eqref{eq:sigmasq}, which
is plotted in Fig~\ref{fig:twotraj2}. It appears that the system
becomes homogeneous in a very short time from the point of view of the
single-particle density operators. Though the monotonicity of the curve is
slightly disturbed by the edge effects due to the finite size of the
system (compare Figs.~\ref{fig:xxprop} and~\ref{fig:twotraj2}), even
these do not significantly alter the behavior.  This kind of coherent
homogenization has a rather clear interpretation: since the initial
``disturbance'' causes a broadening and the state has to remain
normalized, the disturbance is diluted along the chain like water
waves generated by a single drop. Note that since there are many
particles taking part in the process, the homogenization is much
faster than in the case when it is due to pairwise interactions. In
Fig~\ref{fig:twotraj2} we have also plotted the time evolution of the
net concurrence defined in Eq.~\eqref{eq:concsum}. Note that it grows
monotonously till the edge effects appear.

In order to interpolate between the stochastic homogenization and the
coherent one, we consider now a one-dimensional semi-quantal spin gas
model with random classical dynamics.  Let us consider $N$ classical
particles arranged on a chain of length $L$ with periodic boundary
conditions. A configuration of the particles is given by specifying
their position.  The sites are labeled by $l=1\ldots L$. Thus the
specification of a position $l(k)$ for each particle $k=1\ldots N$
defines a configuration of the classical part of the system. Initially
the system is described by a fixed configuration $(l(k))(0)$. The
evolution of the system is stochastic and we consider discrete time
steps $t=0,1,2,\ldots$. The evolution is defined in terms of a
probability transition matrix $P_{(l(k))(t) | (l'(k))(t+1)}$ which
gives the probability that a configuration $l(k)$ at time $t$ evolves
into $l'(k)$ at time $t+1$.  We will consider Markov chains only,
thus the matrix $P$ will be independent of time and the previous
states of the system,
\begin{equation}
  \label{eq:prtran}
 P_{(l(k))(t) |
  (l'(k))(t+1)}=P_{(l(k)) | (l'(k))}
\end{equation}
for all $t$. Actually we will consider the following evolution: at
each time step a random particle and a random direction is chosen. If
the target site is empty, the particle moves there. In this case the
probability transition matrix in Eq.~\eqref{eq:prtran} is symmetric,
thus it describes a doubly stochastic Markov chain. These are known to
have a single stationary probability distribution which is uniform.
Starting the system in an arbitrary configuration, after a time long
enough all the configurations can be found with equal probability. Let
us remark that though the evolution of the system is Markovian, any of
its subsystems evolve in a non-Markovian way as the rest of the system
acts as a memory.

The quantum mechanical part of the system is considered to be in a
pure state $|\Psi_0\rangle$ at $t=0$. In each (discrete) time step
there is an interaction Hamiltonian which depends on the classical
configuration. In the case of the discrete classical model we have
\begin{equation}
  \label{eq:Hgen}
  H(t)=\sum_{<k,k'>_t} H^{(k,k')} \, ,
\end{equation}
where $<k,k'>_t$ means that the particles $k$ and $k'$ are neighbors
at time $t$, while $H^{(k,k')}$ is a chosen bipartite qubit
interaction, in particular the Heisenberg-XX Hamiltonian in
Eq~\eqref{eq:Hxx} in this Section. Note that in this case the
Hamiltonian in Eq.~\eqref{eq:Hgen} is simply twice the adjacency
matrix of the actual configuration of the particles. The evolution is
given by
\begin{equation}
  \label{eq:qevol}
  |\Psi(t+1)\rangle = \exp(-i \eta H(t)) |\Psi(t)\rangle,
\end{equation}
where $\eta$ is the parameter describing the coupling strength. Though
the particular value of $\eta$ affects the fluctuations of the studied
quantities as well as the speed of the transients, it did not affect
the main features of the behaviour in our simulations. Therefore we
present results where $\eta$ has a fixed arbitrary value. However,
special values of $\eta$ for certain Hamiltonians might cause special
behaviour, which would be a possible topic of further studies.  The so
defined evolution, which is the subject of the present study, is a
sequence of unitary evolution steps where the actual unitary step is
determined by the state of the underlying classical Markov chain. 

\begin{figure}
  \centering
   \includegraphics{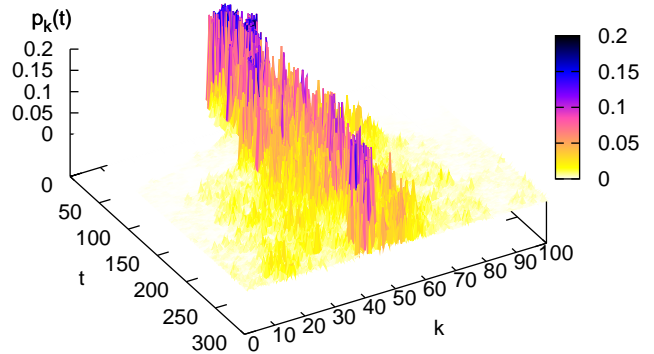}
   \caption{(color online) ``Propagation of the disturbance'' in the
     one-dimensional lattice gas model with 100 particles at 150 sites,
     $\eta=0.1$. The initial state is $|\underline{50}\rangle$, that
     is, all the qubits are in the state $|0\rangle$ except for the
     middle one, which is in the state $|1\rangle$. Note that the probability is plotted against the particle's index $k$. Since the classical evolution preserves the order of the particles, this reflects spatial order of the spins even though their actual position changes. }
  \label{fig:latticegasprop}
\end{figure}

Looking at the evolution of $p_k(t)$ in Eq.~\eqref{eq:rhok} which is
plotted in Fig.~\ref{fig:latticegasprop} we find that there is again a
propagation kind of phenomenon, but it reflects the randomness of the
classical motion.  The disturbance may propagate in the cluster but it
is reflected by the vacancies. These reflections slow down the
propagation and make it rather noisy.
If we consider the evolution of Eq.~\eqref{eq:sigmasq}, which is
plotted in Fig.~\ref{fig:twotraj1} (Upper figure, curves B and C with
different particle densities), we find that even though it shows
significant random oscillations, it has a clear tendency of
homogenization. Also, the behavior of the net concurrence shows a
growing tendency, see Fig.~\ref{fig:twotraj1} (Lower figure, curves B
and C), though it is rather oscillatory and the growth depends highly
on the details of the classical evolution.

\subsection*{The billiard-ball model}

Next we turn our attention to another similar model. Assume that each
of the quantum bits is assigned to a (ball-shaped) particle of a
finite diameter and mass moving classically in three dimensions. For
each evolution, a classical initial condition for the positions
$\mathbf{r}_k$ and momenta $\mathbf{p}_k$ is generated, and the
classical dynamics of the system is solved. The initial positions are
drawn from a uniform (within the three-dimensional region given by the
containing box dimensions reduced by the perimeter of the balls on
each end of the three coordinate intervals) distribution for the ball
origins in three dimensions. Should the generated position of a
particle be such that the ball representing the particle overlaps with
any of previously generated ones, the position is discarded and a new
one is generated. The distribution of momenta is Boltzmann with zero
mean and a fixed standard deviation. Two particles (or a particle and
the containing box) collide elastically whenever a contact of their
boundaries takes place. The collisions (i.e.~the appropriate changes
of the colliding particles' velocities) are instantaneous. The mass of
the containing box is infinite, i.e.~its position does not vary
upon particle-box collisions.

The quantum system is initially in the state in Eq.~\eqref{eq:psiinsubs_normalbase} again. During
the evolution a pair of qubits interacts via the gate in
Eq.~\eqref{eq:xxgate} iff the two respective billiard balls collide.
This is a kind of semi-quantal spin gas model.

\begin{figure}
  \centering
  \includegraphics{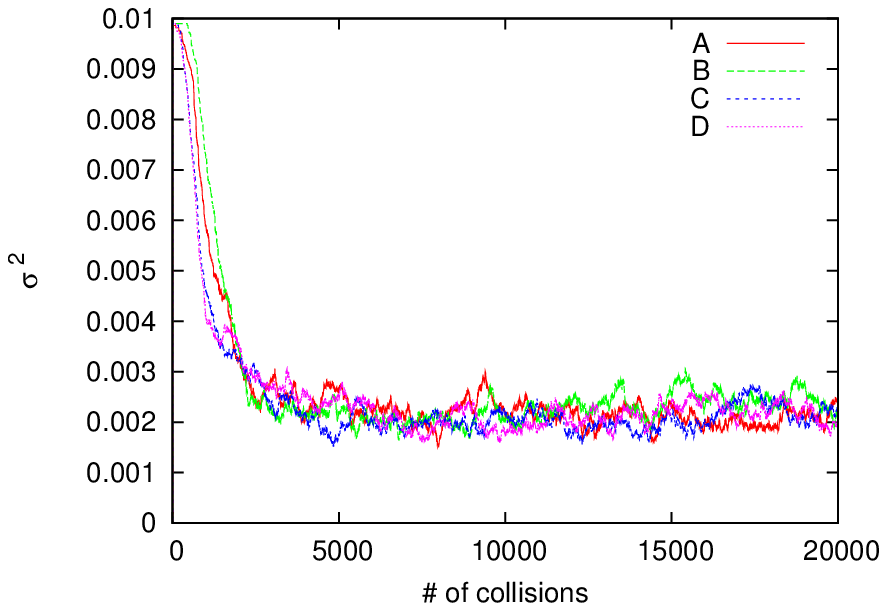}

  \includegraphics{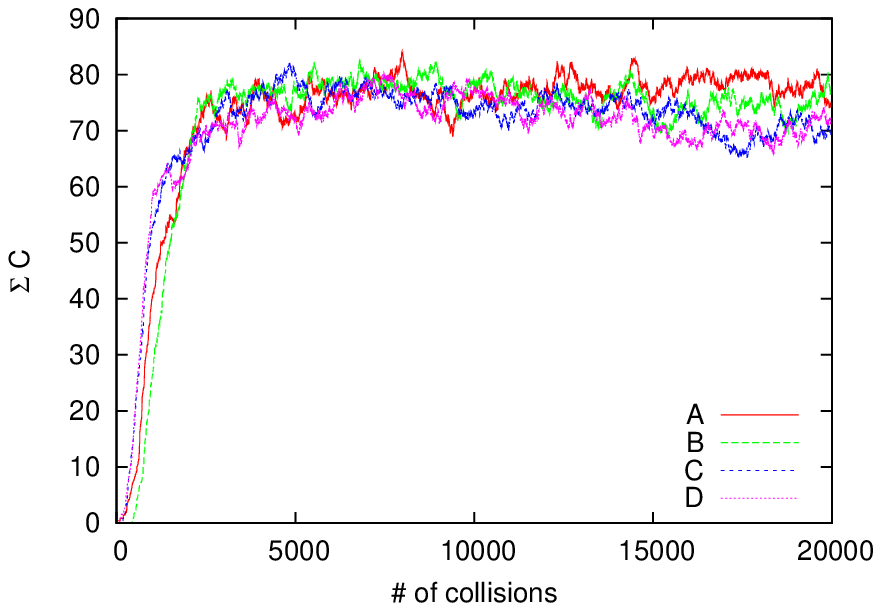}
  \caption{(color online) Evolution of the degree of inhomogeneity in
    Eq.~\eqref{eq:sigmasq} (upper figure) and the sum of the pairwise
    concurrences in Eq.~\eqref{eq:concsum} (lower figure) for
    two billiard-ball collision sequences (A, B) and for two random collision sequences (C, D).
    $N=100$ ball-shaped particles of diameter
  one (arbitrary units) in a cube box with dimensions
  $150\times150\times150$ (arbitrary units). The mass of the
  particles is one (arbitrary units). The initial velocities (and hence also
  the velocities at later times) are normally distributed with zero mean and standard deviation $\sigma=0.32$ (arbitrary units). The interaction
    strength parameter in Eq.~\eqref{eq:xxgate} and
    Eq.~\eqref{eq:xxchainu} is $\eta=0.1$.
    }
  \label{fig:billiardballs}
\end{figure}

We again look at whether homogenization features are present in this model and find qualitatively the same behavior as in the case of the spin chain with vacancies model. Sequences generated by gas dynamics and random ones yield very similar results (see Fig.~\ref{fig:billiardballs}).

We can conclude that quantum homogenization at a single particle level
appears in a variety of models based on the same bipartite interaction,
albeit with very different underlying classical dynamics. These models
range from the random bipartite interactions where each particle can
interact directly to one-dimensional diffusive lattice gases or regular
spin chain where there is no direct interaction of the particles.

\section{State randomization due to stochastic classical dynamics}
\label{sect:globalstate}

In this Section we consider the diffusive lattice gas model already
described, and investigate the effect of the loss of information on
the stochastic evolution of the classical system.

We simulate the evolution of the system on a computer with a fixed
classical and quantum initial condition. Unlike in the previous
Section, we run the simulation now $s$ times independently, which
yields different $|\Psi_s(t)\rangle$ evolutions, depending on the
details of the evolution on the classical subsystem.  We will refer
to this as the $s$-th trajectory.  Simulating $N_{\text{traj}}$ trajectories we
construct the density operator
\begin{equation}
  \label{eq:rhotaprox}
  \varrho(t)=
  \frac1{N_{\text{traj}}} \sum_{s=1}^{N_{\text{traj}}}
  | \Psi_s(t)\rangle\langle  \Psi_s(t)|.
\end{equation}
Its interpretation is as follows: if we disregard all the information
we have on the classical evolution of the system we can describe its
state by a density operator
\begin{equation}
  \label{eq:varrhot}
  \tilde{\varrho}(t)=\sum_{s'} p_{s'}(t) |\Psi_{s'}(t)\rangle\langle \Psi_{s'}(t)|,
\end{equation}
where $|\Psi_{s'}(t)\rangle$-s are all the possible evolutions of the
system and $p_{s'}(t)$ is the probability of obtaining the given
evolution. We expect that for $N_{\text{traj}}\to \infty$, $\varrho(t)$ will
converge to the exact $\tilde{\varrho}(t)$ stochastically. In
practice we determine the required number of simulations $N_{\text{traj}}$
empirically by increasing it till the further increase does not
decrease the variance of the density operator elements.
The evaluation of~\eqref{eq:rhotaprox} is obviously suitable for
parallel computing. Thus we have carried out our simulations on a
parallel computer using GNU Octave and its MPI toolbox~\cite{mpitb}.

We consider the evolution of the von Neumann entropy of the whole
system,
\begin{equation}
  \label{eq:neumannent}
  S(t)=-\mathop{\mbox{Tr}}\varrho(t)\log_2
  \varrho(t)
\end{equation}
which describes its mixedness. In addition, we will consider bipartite
entanglement as measured by concurrence in Eq.~\eqref{eq:concurrence}

\begin{figure}
  \centering
  \includegraphics{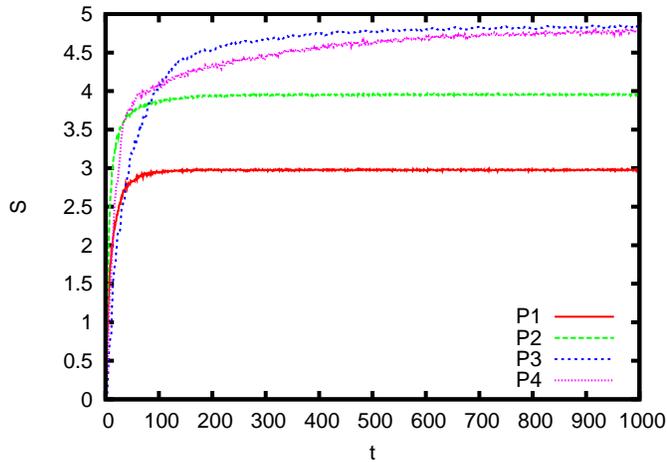}
  \caption{(color online) Time evolution of the von Neumann entropy of the density
    operator in a Heisenberg-XX spin gas with various parameters.  The
    interaction strength is $\eta=1.0$ for all the curves, 3000
    trajectories were simulated in each case.  P1: 8 particles, 16
    sites; P2: 16 particles, 20 sites; P3: 32 particles, 33 sites; P4:
    32 particles, 40 sites.}
  \label{fig:xxpurity}
\end{figure}

\begin{figure}
  \centering
  \includegraphics{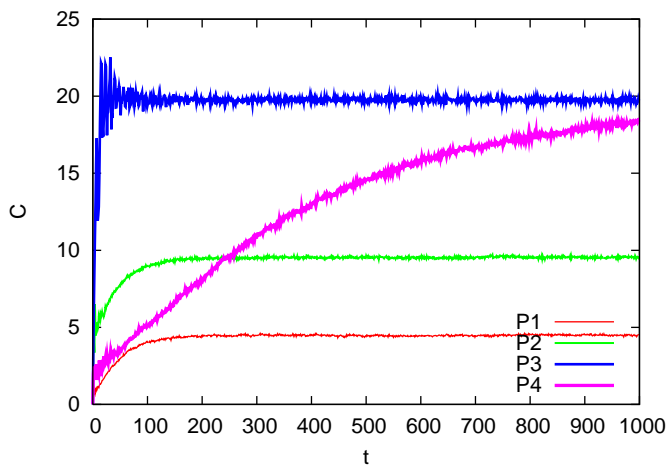}
  \caption{(color online) The average concurrence in
    Eq.~\eqref{eq:netentavg} in a one-dimensional lattice gas with
    Heisenberg XX interactions.   The
    interaction strength is $\eta=1.0$ for all the curves, 3000
    trajectories were simulated in each case.  P1: 8 particles, 16
    sites; P2: 16 particles, 20 sites; P3: 32 particles, 33 sites; P4:
    32 particles, 40 sites.}
\label{fig:concavg}
\end{figure}

\begin{figure}
  \centering
  \includegraphics{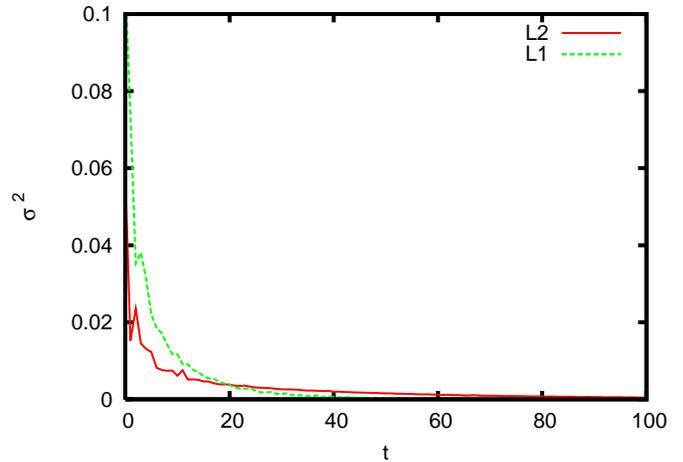}
  \caption{(color online) The inhomogeneity in Eq.~\eqref{eq:sigmasq} of the density
    operator of the system calculated according to
    Eq.~\eqref{eq:rhotaprox}, in a one-dimensional lattice gas with
    Heisenberg-XX interactions. There were 3000 trajectories simulated
    to obtain the density operator. The coupling strength was
    $\eta=1$. Parameters of the curves: L1: 8 particles, L2=16
    particles, 40 sites in both cases.}
  \label{fig:rhothomog}
\end{figure}

First we consider the $XX$ Hamiltonian in Eq.~\eqref{eq:Hxx} as the
bipartite interaction. As all the Hamiltonians in Eq.~\eqref{eq:Hgen}
commute with the total $z$ component of the spin, the subspace spanned
by the vectors in Eq.~\eqref{eq:subspacebasis} will be invariant in
this case too. Consider the state $|\underline{1}\rangle$ as an
initial state. In Fig.~\ref{fig:xxpurity} we plot the time evolution
of the von Neumann entropy of the state of the whole system as defined
in Eq.~\eqref{eq:rhotaprox}, i.e. the density operator in the absence
of information on the collision history. It appears that after some
transient which depends on the features of the underlying dynamics
(e.g. its entropy rate), the system will be in the complete mixture
of the given subspace,
\begin{equation}
  \label{eq:subspacemix}
  \varrho_{\text{mix}}=\frac1N \sum_{k=1}^N |\underline{k}\rangle\langle \underline{k}|,
\end{equation}
which is reflected in the maximum value of von Neumann entropy, i.e.
$\log_2 N$. Note that this already appears if there is a single
unoccupied site in the system, thus the effect of the classical noise
is rather remarkable.
Thus the situation of the microcanonical ensemble is arising in this
model: the state of the system is an equal-weight mixture of all of
its possible states. This implies that there is a vanishing stationary
entanglement, of course. If, however, we calculate the average net
concurrence for each trajectory and calculate the average afterwards,
\begin{equation}
  \label{eq:netentavg}
  \overline{C}(t)= \frac1{N_{\text{traj}}} \sum_{s=1}^{N_{\text{traj}}}
  C_{\text{tot}}(|\Psi_s(t)\rangle),
\end{equation}
(C.f. Eq.~\eqref{eq:concsum}), we find that there is a nonvanishing
concurrence, see Fig.~\ref{fig:concavg}.
This is in accordance with our expectations: as we have found in the
previous Section, homogenization takes place in each
individual trajectory which leads to a nonvanishing concurrence. If,
however, we drop the information on the classical evolution, this
resource will not be accessible. Finally, it is also interesting to
plot the degree of inhomogeneity in Eq.~\eqref{eq:sigmasq} of the
density operator of the system calculated according to
Eq.~\eqref{eq:rhotaprox}. This is done in Fig.~\ref{fig:rhothomog}. It
appears that the homogenization becomes smooth due to the averaging
over several trajectories.
It is interesting to point out here that if we consider an initial
state which is a superposition of two states from different invariant
subspaces, the stationary entropy will be initial state dependent. As
we have verified via simulations as well, if one considers an initial state
\begin{equation}
  \label{eq:psiintwosubs}
  |\Psi(t=0)\rangle=C_0|\underline 0\rangle + C_1 |\underline{1}\rangle,
\end{equation}
where $|\underline{0}\rangle= |0\rangle^{\otimes N}$, the stationary
entropy will be
\begin{equation}
  \label{eq:sinitialstate}
  S_{\text{stationary}}=
  H\left(|C_0|^2,\frac{|C_1|^2}N, \frac{|C_1|^2}N,\ldots \frac{|C_1|^2}N\right),
\end{equation}
where $H()$ is the Shannon entropy function, and the argument
$\frac{|C_1|^2}N$ appears $N$ times. This rather expectable effect is
very quantum mechanical: the conservation laws produce invariant
subspaces and the stationary entropy of the microcanonical
distribution depends on the relation of a given initial state to this
subspace structure. The situation is conceptionally similar to the
phase-like transitions in the Dicke model~\cite{buzek:163601}.

Finally, let us first consider the Ising interaction
\begin{equation}
  \label{eq:isingx}
  H_{\text{Ising}}=\sigma_x \otimes \sigma_x,
\end{equation}
as the bipartite interaction for Eqs.~\eqref{eq:Hgen}
and~\eqref{eq:qevol}. It would dynamically generate cluster states
from the products of the eigenstates of the $\sigma_z$ operators in a
usual spin chain (i.e., with no empty sites or classical motion).  The
quantum system is the product state $|100\ldots0\rangle$ initially,
which would evolve into a cluster state if the lattice was
saturated. In this case the possibility of numerical simulation is
limited by the exponentially growing size of the Hilbert
space. However, it appears according to the simulations that the
entropy tends to converge to $N-1$ in each case. Looking at the
density matrix in the computational basis it will be diagonal and it
describes a complete mixture in a subspace spanned by the vectors of
the computational basis with odd Hamming weights, which has a
dimensionality half of the whole Hilbert space. It is easy to see that
the subspaces spanned by the computational basis vectors with even or
odd Hamming weights are invariant subspaces of the pairwise Ising
interactions.
\begin{figure}
  \centering
  \includegraphics{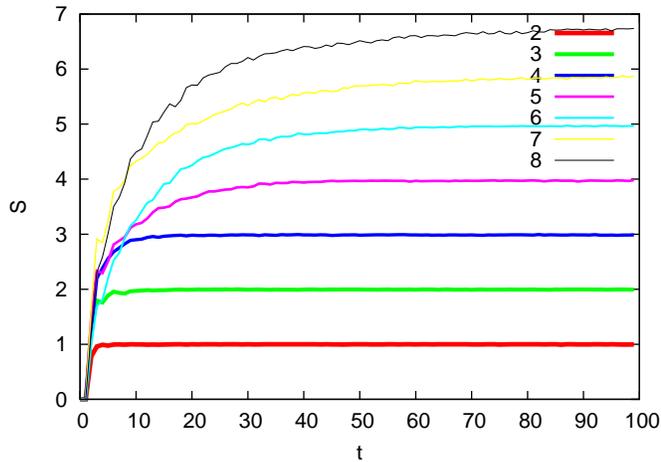}
  \caption{(color online) Evolution of the entropy of the state of an
    Ising spin gas for the quantum initial state $|100\ldots0\rangle$,
    with a fixed random uniform classical initial condition. The
    interaction strength $\eta$ is set to $1$. The parameter of the
    curves is $N$, the number of particles, the length of the chain is
    chosen to be $L=3N$ with periodic boundary conditions. 3000
    trajectories were simulated to obtain the figure.}
  \label{fig:isingmult}
\end{figure}

\begin{figure}
  \centering
  \includegraphics{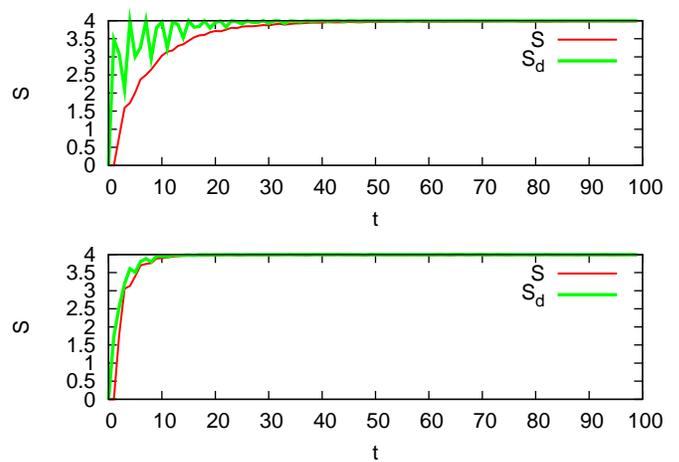}
  \caption{(color online) Evolution of the von Neumann entropy of the
    state ($S$) and the Shannon-entropy of the diagonal of the density
    operator ($S_d$) in the computational basis of an Ising spin gas
    for the quantum initial state $|100\ldots0\rangle$, with a fixed
    random uniform classical initial condition. In the upper figure
    there are $5$ particles on $6$ sites considered, while in the
    lower figure the $5$ particles are located at a chain of $10$
    sites.  The interaction strength $\eta$ is set to $1$. 3000
    trajectories were simulated to obtain the figure.}
  \label{fig:isingrand}
\end{figure}

It is also interesting to compare the (classical Shannon) entropy of
the probability distribution in the diagonal of the density matrix in
the computational basis with the von Neumann entropy of the
density operator. Since this latter is the infimum of the entropies of
the diagonal of the density matrices taken over all the possible
bases, if these two are close to each other the density matrix is
almost diagonal. The evolution of these entropies is plotted in
Fig.~\ref{fig:isingrand}.
Obviously since the density operator tends to a complete mixture in
a subspace, the stationary values will be the same. If there are just
a few vacancies in the system, the two entropies differ. If, however,
there is a relevant number of unoccupied sites in the system, one
finds that the density operator during the whole evolution will be
``almost diagonal'' in the computational basis in the sense that the
entropy of the diagonal of the density matrix will be almost the same
as the von Neumann entropy of the density operator. Let us remark that
the same can be observed for the Heisenberg-XX interaction in the
basis in Eq.~\eqref{eq:subspacebasis}.  Since these bases are built up
entirely from product states, one can conclude that the evolution for
a large number of empty sites is very similar to a classical
stochastic process: the arising decoherence is very fast.

All together we find that in spin gas models with random classical
dynamics the details of the classical motion are required to access
quantum features of the system. In the absence of these the system
will attain a highly mixed stationary state, which is essentially
independent of the classical motion. The transients, however, depend
on these details, such as the entropy rate of the classical Markov
process, rules governing the interaction, etc.

\section{Summary and conclusions}
\label{sect:concl}

We have investigated various semi-quantal spin systems as microscopic
models of decoherence. One of the aspects was to study the presence of
quantum homogenization at the single particle level, purely due to the
interaction between the qubits in argument. We have found that with
Heisenberg-XX couplings homogenization always appears, accompanied by
a stationary net entanglement, regardless of the details of the
evolution of the classical part of the system.

We have also studied the mixedness of the whole system in the absence
of any information on the particular classical evolution. We have
found that at least in the studied cases the system will reach a
stationary state which is a microcanonical ensemble in the quantum
mechanical part.  This in turn implies that the system qubit will
finally decohere, too.  Similarly to classical non-periodic Markov
chains the transients leading to this stationary state depend on the
details of the underlying quantum dynamics, while the stationary state is
the same.  This is true even for classical processes with a small
entropy rate.

\acknowledgements

M. K. acknowledges the support of the Hungarian Scientific Research
Fund (OTKA) under the contract No. T049234. This work was supported in
part by the European Union projects CONQUEST and QAP, by the Slovak
Academy of Sciences via the project CE-PI/2/2005, by the project
APVT-99-012304. The numerical computations were carried out on the HPC
facility of Faculty of Science, University of P\'ecs.

%

\begin{thebibliography}{10}

\bibitem{RossiniCGMF07}
D. Rossini {\it et~al.}, J. Phys. A: Math. Gen. {\bf 40},  8033  (2007).

\bibitem{TessieriW03}
L. Tessieri and J. Wilkie, J. Phys. A: Math. Gen. {\bf 36},  12305  (2003).

\bibitem{DawsonHMM05}
C.~M. Dawson, A.~P. Hines, R.~H. McKenzie, and G.~J. Milburn, Phys. Rev. A {\bf
  71},  052321  (2005).

\bibitem{PhysRevA.61.052306}
V. Coffman, J. Kundu, and W.~K. Wootters, Phys. Rev. A {\bf 61},  052306
  (2000).

\bibitem{osborne:220503}
T.~J. Osborne and F. Verstraete, Phys. Rev. Lett. {\bf 96},  220503  (2006).

\bibitem{CamaletC07}
S. Camalet and R. Chitra, Phys. Rev. B {\bf 75},  094434  (2007).

\bibitem{PinedaGS07}
C. Pi{\~ n}eda, T. Gorin, and T.~H. Seligman, New J. Phys. {\bf 9},  106
  (2007).

\bibitem{ZimanSBHSG02}
M. Ziman {\it et~al.}, Phys. Rev. A {\bf 65},  042105  (2002).

\bibitem{ScaraniZSGB02}
V. Scarani {\it et~al.}, Phys. Rev. Lett. {\bf 88},  097905  (2002).

\bibitem{DiosiFK06}
L. Di{\' o}si, T. Feldmann, and R. Kosloff, Int. J. Quant.
  Inf. {\bf 4},  99  (2006).

\bibitem{ZimanSB03}
M. Ziman, P. {\v S}telmachovi{\v c}, and V. Bu{\v z}ek, J. Opt. B-Quantum
  Semicl. Opt. {\bf 5},  S439  (2003).

\bibitem{BenentiP07}
G. Benenti and G.~M. Palma, Phys. Rev. A {\bf 75},  052110  (2007).

\bibitem{HartmannCDB05}
L. Hartmann, J. Calsamiglia, W. Dur, and H.~J. Briegel, Phys. Rev. A {\bf 72},
  052107  (2005).

\bibitem{christandl:187902}
M. Christandl, N. Datta, A. Ekert, and A.~J. Landahl, Phys. Rev. Lett. {\bf
  92},  187902  (2004).

\bibitem{KoniorczykRB05}
M. Koniorczyk, P. Rap{\v c}an, and V. Bu{\v z}ek, Phys. Rev. A {\bf 72},
  022321  (2005).

\bibitem{mpitb}
M. Creel, Computational Economics {\bf 26},  107  (2005).

\bibitem{buzek:163601}
V. Bu{\v z}ek, M. Orszag, and M. Ro{\v s}ko, Phys. Rev. Lett. {\bf 94},  163601
  (2005).

\end{thebibliography}

\end{document}